\documentclass[aps,pra,a4paper, twocolumn, 10pt]{revtex4-1} 
\usepackage{bbm, amsmath, amssymb, amsthm, bm, textcomp,graphicx,color}
\usepackage[utf8]{inputenc}
\usepackage[T1]{fontenc}
\usepackage[english]{babel}

\theoremstyle{definition}

\begin{document}
\title{Lower bounds on the size of Schr\"odinger's cat from experimental data}
\author{Florian Fr\"owis}
\affiliation{Group of Applied Physics, University of Geneva, 1211 Geneva, Switzerland}
\date{\today}
\begin{abstract}
  Experimental progress with meso- and macroscopic quantum states (i.e., general Schr\"odinger-cat states) was recently accompanied by theoretical proposals on how to measure the merit of these efforts. So far, experiment and theory were disconnected as theoretical analysis of actual experimental data was missing. Here, we consider a proposal for macroscopic quantum states that measures the extent of quantum coherence present in the system. For this, the quantum Fisher information is used. We calculate lower bounds from real experimental data. The results are expressed as an ``effective size'', that is, relative to ``classical'' reference states. We find remarkable numbers of up to 70 in photonic and atomic systems. 
\end{abstract}

\maketitle

\section{Introduction}
\label{sec:introduction}

Quantum experiments involving many particles, modes or excitations do not only require a large experimental efforts. These experiments are also challenging from a theoretical point of view. Theory has to provide a meaningful and accessible characterization that also allows one to compare different physical systems or quantum state classes. Just giving the full information (i.e., state tomography) is not only technically tough but is generally hard to process and assess. A candidate for a key property of large systems is the quantum Fisher information (QFI). A large QFI implies usefulness in quantum metrology, multipartite entanglement in composed systems and rapid state evolution (see \cite{Toth_Quantum_2014} for an overview including references).

In this paper, we show how to extract lower bounds on the QFI from real experimental data and present explicit analysis for several experiments with photons and atoms. While this analysis is independent of the application of the QFI, here we emphasis the role of the QFI in the context of macroscopic quantum states as a generalization of Schr\"odinger-cat state. As detailed below, a large QFI is argued to be a characteristic trait of superpositions of macroscopically distinct states \cite{Frowis_Measures_2012}. By comparing the QFI of the actual experiment to a ``classical'' reference state (coherent state or product state) we define a so-called effective size (or simply size), which tells us ``how much larger'' the quantum state is compared to a microscopic unit. By applying lower bounds on recent experimental data from \cite{Eberle_Stable_2013,Hosten_Measurement_2016,Monz_14-Qubit_2011,Vlastakis_Deterministically_2013,Kienzler_Observation_2016,Wang_Schrodinger_2016,Vahlbruch_Detection_2016}, we show that with modern experiments sizes up to 70 are reachable.

The paper is structured as follows. In the next section, we review the motivation for choosing the QFI as a measure for macroscopic quantum states as it detects the extent of quantum coherence in well-chosen spectra. In section~\ref{sec:lower-bounds-qfi} we discuss few properties of the QFI that lead to accessible lower bounds on the QFI. These bounds are applied to real experimental data in section~\ref{sec:appl-exper}. The impact of error bars and the data fitting is analyzed. Finally, a few additional aspects are discussed in section~\ref{sec:disc-concl}.

\section{The QFI as a measure for macroscopic quantum states}
\label{sec:quant-fish-inform}

Recently, there have been some theoretical effort in characterizing the main properties of quantum states of meso- or macroscopic systems that reflect Schr\"odinger's gedankenexperiment of a cat in superposition of being dead and alive. Although there is no commonly accepted single measure of macroscopic quantum states, one can identify a few similarities in some proposals. 
More specifically, many contributions \cite{Leggett_Macroscopic_1980,Shimizu_Stability_2002,Bjork_size_2004,Cavalcanti_Signatures_2006,Korsbakken_Measurement-based_2007,Marquardt_Measuring_2008,Lee_Quantification_2011,Frowis_Measures_2012,Sekatski_Size_2014,Laghaout_Assessments_2015} explicitly or implicitly agree that a macroscopic quantum state should show large quantum coherence (or large quantum fluctuations) spread over a ``reasonably chosen'' spectrum. That is, a macroscopic quantum  state should be in superposition of states that are far apart in the spectrum (like the biological cat being in a superposition of two distant parts in some kind of ``vitality'' spectrum). For pure states, one way to argue is the following \cite{Shimizu_Stability_2002}. Consider a large, isolated system composed of $N \gg 1$ microscopic constituents (``particles'') and the chosen spectrum comes from $X = \sum_i x^{(i)}$ as the sum of bounded microscopic observables ($X$ is called a collective observable). We are already convinced that quantum mechanics holds for single particles and therefore quantum coherence $\Delta x$ is possible on this level (up to to the order of the spectral radius of $x$). However, if no quantum correlation between the particles is allowed, the quantum coherence on the macroscopic scale is just additive $\Delta X = \sqrt{N} \Delta x$. Compared to the spectral radius $\varrho \geq N \Delta x$, this coherence is relatively small since $\Delta X/\varrho \leq N^{-1/2}$. Hence, the observation of persistent quantum coherence on the macroscopic level is sufficient to show strong quantum correlation between particles, which is sometimes called a ``macroscopic quantum effect''. 

The choice of the spectrum is as crucial as the choice of the partition of a Hilbert space into subspaces in entanglement theory. Like in the latter, one can argue for some natural choices. In particular, observables that have a macroscopic limit (e.g., magnetization, electric fields or position) are considered as candidates. Typically, there are several observables with this property for a given system. The set of these observables is denoted as $\mathcal{X}$.

After defining $\mathcal{X}$, a simple and intuitive way to measure the spread of the coherence for pure states is to calculate the variance $(\Delta X)^2_{\psi}$ for given state $| \psi \rangle $ and observable $X \in \mathcal{X}$ (see reference \cite{Yadin_general_2016} for a rigorous information theoretic argument in favor of the variance). In the general case, a large variance might also come from incoherent mixtures of pure states. A way to distinguish between coherent and incoherent parts in the variance is to construct the convex roof of the variance. This is a well-established concept in quantum information theory and reflects a kind of ``worst-case'' scenario. Given all pure state decompositions (PSD) of a quantum state $\rho = \sum_n p_n \left| \psi_n \right\rangle\!\left\langle \psi_n\right| $ (where the $| \psi_n \rangle $ are not necessarily orthogonal), one looks for the decomposition that minimizes the average variance 
\begin{equation}
\label{eq:1}
\mathcal{I}_{\rho}(X) = \min_{\mathrm{PSD}} \sum_{n} p_n (\Delta X)^2_{\psi_n}.
\end{equation}
As it was shown in reference \cite{Yu_Quantum_2013}, the QFI is four times the convex roof of the variance. In the following, we neglect the factor four and directly call $\mathcal{I}$ the QFI. Hence, the QFI measures the quantum coherence of $\rho$ in the spectrum of $X$ and is therefore a notion of coherence length in $X$.

To have a normalized measure, one defines ``classical states'', that is, pure quantum states $\psi_{\mathrm{cl}} = | \psi_{\mathrm{cl}} \rangle \! \left\langle \psi_{\mathrm{cl}} \right| $ that have a minimal variance with respect to all $X \in \mathcal{X}$. Although one has to take care of some subtleties, it is often easy to identify all $\psi_{\mathrm{cl}}$. Denoting the set of classical states by $\mathcal{C}$, one defines the effective size $N_{\mathrm{eff}}(\rho)$ of a quantum state $\rho$ as
\begin{equation}
\label{eq:2}
N_{\mathrm{eff}}(\rho) = \max_{X \in \mathcal{X}} \frac{\mathcal{I}_{\rho}(X)}{\max_{\psi_{\mathrm{cl}} \in \mathcal{C}}\mathcal{I}_{\psi_{\mathrm{cl}}}(X)}.
\end{equation}
This number tells us how much larger the quantum coherence of $\rho$ is compared to the maximal coherence generated by classical states $\psi_{\mathrm{cl}}$.

To be more specific, let us consider two relevant examples. The first one is the phase space with the canonical relation $[a,a^{\dagger}] = \mathbbm{1}$. If we define the quadrature operators $X_{\vartheta} = e^{i \vartheta} a + e^{-i \vartheta} a^{\dagger}$ as the set $\mathcal{X}$, the coherent state $| \alpha \rangle $ (defined via $a \left| \alpha \right\rangle = \alpha \left| \alpha \right\rangle $) is the most classical state, which coincides with the long history of coherent states used to describe classical states of light. Since $\mathcal{I}_{\alpha}(X_{\vartheta}) = 1$ for all $\vartheta$, the effective size for single-mode states reads $N_{\mathrm{eff}}(\rho) = \max_{\vartheta} \mathcal{I}_{\rho}(X_{\vartheta})$. For systems with $N$ modes, $\mathcal{X}$ is extended to sums of quadratures of the individual modes $X_{\vec{\vartheta}} = \sum_{i=1}^{N} X^i_{\vartheta_i} $. The normalization factor from products of coherent state is $N$ (see reference \cite{Oudot_Two-mode_2015} for details).
As a second example, consider an ensemble of $N$ two-level systems. A canonical choice of $\mathcal{X}$ is the set of all operators that are sums of single-particle operators (with unit operator norm, see \cite{Frowis_Measures_2012}). In the most general case, one has to optimize over two angles per particle, $\varphi_i$ and $\vartheta_i$, but symmetries often help to reduce the complexity. Classical states in our sense are then product state that exhibit a maximal variance equaling $N$. Thus, for both cases the effective size reads 
\begin{equation}
\label{eq:4}
N_{\mathrm{eff}}(\rho) = \frac{1}{N} \max_{X \in \mathcal{X}} \mathcal{I}_{\rho}(X).
\end{equation}
The effective size is tailored such that the maximal value is proportional to the total number of particles or excitations. A few important examples for the phase space and spin ensembles are listed in tables \ref{tab:purePS} and \ref{tab:pureN}, respectively.

\begin{table}[hb]
  \centering\setlength{\tabcolsep}{7pt}
  \begin{tabular*}{\columnwidth}{l l l}
      \hline  \hline
      State & $N_{\mathrm{eff}}$ & Remark\\
    
      \hline
      Squeezed state $\left| S_r \right\rangle $ & $\exp(2r)$ & one and two modes\\
      Fock state $| n \rangle $ & $2n+1$ & \\
      Cat state $ \left| \alpha \right\rangle + \left| -\alpha \right\rangle $ & $\approx 4 |\alpha|^2 + 1$ & valid for $|\alpha| \gtrsim 1$\\
      \hline  
    \end{tabular*}
  \caption{Examples for states with large $N_{\mathrm{eff}}$ in phase space.}
  \label{tab:purePS}
\end{table}

\begin{table}[hb]
  \begin{tabular*}{\columnwidth}{l l l}
    \hline  \hline
    State & $N_{\mathrm{eff}}$ & Remark\\
\hline
    Spin-squeezed state & $O(N^{2/3})$ & one-axis twisting \cite{Kitagawa_Squeezed_1993}\\
    Oversqueezing & $O(N)$ &  one-axis twisting \cite{Pezze_Entanglement_2009}\\
    Dicke state $| N,k \rangle $ & $\frac{2k(N-k)}{N-1} + 1$ & $k$ excitations\\
    GHZ state $ \left| 0 \right\rangle^{\otimes N} + \left| 1 \right\rangle^{\otimes N} $ & $N$ & maximal effective size\\
    \hline
  \end{tabular*}
  \caption{Examples for ensembles of $N$ two-level systems with large $N_{\mathrm{eff}}$.}
  \label{tab:pureN}
\end{table}

To summarize, equation (\ref{eq:2}) [with the example equation (\ref{eq:4})] measures the relative quantum coherence (called effective size) of a quantum state after we have defined a set of macroscopic observables $\mathcal{X}$ and classical states $\mathcal{C}$.

\section{Lower bounds on the QFI}
\label{sec:lower-bounds-qfi}

To be  useful, definitions for macroscopic quantum states have to be applicable to experiments. The QFI-based measure has this property. Despite the fact that $\mathcal{I}_{\rho}(X)$ is generally only computable given the spectral decomposition of $\rho$, there exist tight and accessible lower bounds. To derive the bounds, we note that  $\mathcal{I}_{\rho}(X)$ is intimately connected to how much $\rho$ changes under small perturbations $U = \exp(-i X  \theta)$ (small here means that $\theta \sqrt{\mathcal{I}_{\rho}(X)} \ll 1$), which is the reason why the QFI is important in metrology. Based on this observation, one can derive static and dynamic bounds.

The static variant is a tighter version of the well-known Heisenberg-Robertson uncertainty relation \cite{Kholevo_Generalization_1974,Hotta_Quantum_2004,Pezze_Entanglement_2009,Frowis_Tighter_2015} 
\begin{equation}
\label{eq:5}
\mathcal{I}_{\rho}(X) \geq \frac{\langle i [X,Y] \rangle_{\rho}^2}{4(\Delta Y)_{\rho}^2},
\end{equation}
which holds for any pair of observables $X,Y$. Hence, by measuring the variance of $Y$ and the expectation value of $Z = i[X,Y]$, one directly bound $\mathcal{I}_{\rho}(X)$ from below. Equation (\ref{eq:5}) is particularly useful for squeezed states, where $Y$ corresponds to the squeezed quadrature and $X$ to the antisqueezed one. In phase space, $Z$ is the identity and the bound is tight for all single-mode Gaussian states \cite{Frowis_Tighter_2015}. For spin ensembles, $Z$ is the axis of polarization and similar statements hold. Theoretically, other states with large QFI such as Dicke states and so-called over-squeezed states potentially have tight bounds, but the operators to measure become more cumbersome \cite{Frowis_Tighter_2015}. 

The second approach is to directly witness the effect of $U$. As detailed in reference \cite{Frowis_Detecting_2016}, measurements before and after the application of $U$ give rise to a lower bound on the QFI (see references \cite{Strobel_Fisher_2014,Pezze_Witnessing_2015,Macri_Loschmidt_2016} for similar approaches). For any fixed measurement, let $p_i$ and $q_i$ denote the probability distributions arising from measuring $\rho$ and $U \rho U^{\dagger}$, respectively. Then, the Bhattacharyya coefficient $B = \sum_i \sqrt{p_i q_i}$ is used to lower bound the QFI via
\begin{equation}
\label{eq:6}
\mathcal{I}_{\rho}(X) \geq \frac{1}{\theta^2} \arccos^2 B.
\end{equation}
Effectively, this bound corresponds to a simulation of an estimation protocol where the parameter $\theta$ is controlled and known. There is a connection between large QFI witnessed via equation (\ref{eq:6}) and the violation of Leggett-Garg inequalities \cite{Knee_strict_2016} with ``hardly-invasive'' measurements \cite{Frowis_Detecting_2016}.

It turns out that, for some important instances, the (quasi-)optimal measurement for equation (\ref{eq:6}) is simple, at least for moderate system sizes. As an example, let us discuss the superposition of two coherent states $\left| \mathrm{cat} \right\rangle \propto | \alpha \rangle + e^{i \phi} \left| -\alpha \right\rangle$ (simply called cat state in the following). The maximal QFI reads $\mathcal{I}_{\mathrm{cat}}(x) \approx S + 1$ with $S = 4|\alpha|^2$ (see table \ref{tab:purePS}). The optimal $X_{\theta}$ has the same phase as $\alpha$ (which is taken to be real in the following). The unitary operator $U$ is hence a displacement in phase space.
Experimentally, this is a well-established operation in many setups and the first step to measure the Wigner function $W(x,y)$. After applying $U$, one measures the parity $\Pi$ in the Fock basis. Up to a normalization, a cut of the Wigner function reads $W(0,\theta) =\mathrm{Tr} \Pi U \rho U^{\dagger}$. For the cat state example, one has 
\begin{equation}
\label{eq:7}
W_{\mathrm{cat}}(0,\theta) = A e^{-2\theta^2} \cos(2 \sqrt{S} \theta + \phi),
\end{equation} where $A = 1$ in the ideal case \cite{Raimond_Exploring_2006}. With $p_{\pm} = 1/2[1\pm W(0,\theta_1)]$ and $q_{\pm} = 1/2[1\pm W(0,\theta_2)]$, bound (\ref{eq:6}) is tight for $\phi = 0$ and in the limit $\theta_1 = 0 $ and $\theta_2 \rightarrow 0$.

Note that similar techniques can be applied to atomic ensemble (see, e.g., reference \cite{Monz_14-Qubit_2011} for trapped ions). There, the displacement is replaced by a collective rotation, while the measurement is the parity of the population. From the response of the system to different rotation axes, one finds lower bounds on the QFI. 

\section{Application to experiments}
\label{sec:appl-exper}

\begin{figure}[htbp]
\centerline{\includegraphics[width=1.15\columnwidth]{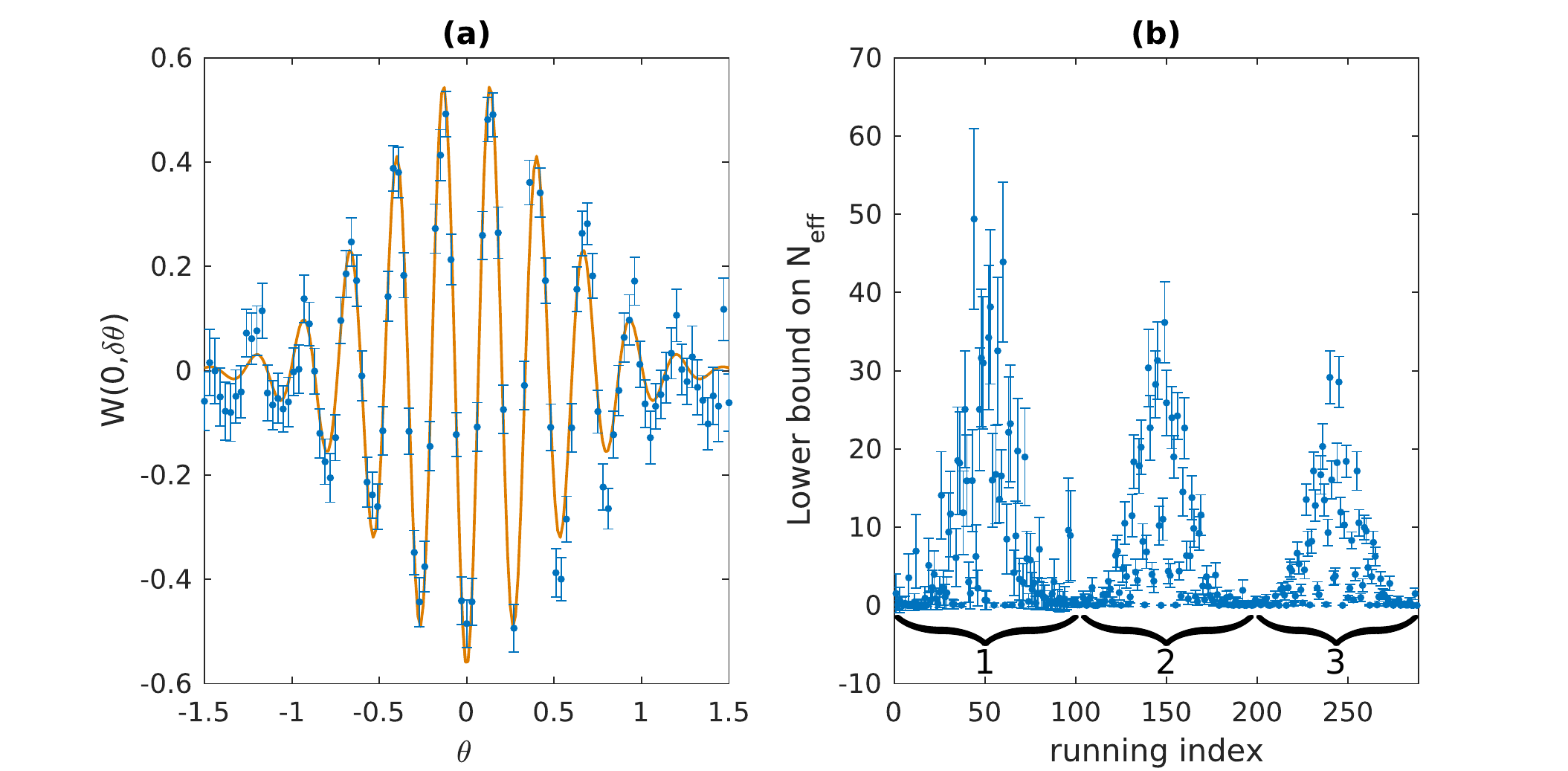}}
\caption[]{\label{fig:home} (a) Data and fit for the Wigner function cut of the cat state with $\alpha = 5.9$ generated in reference \cite{Kienzler_Observation_2016}. (b) Application of bound (\ref{eq:6}) to pairs of data points from (a) with distance one, two and three in units of an elementary step in $\theta$. Pairs from the middle of the data set give better bounds due to the Gaussian envelope.}
\end{figure}

\begin{table*}[htbp]
  \centering\setlength{\tabcolsep}{10pt}
  \begin{tabular*}{0.7\textwidth}{l l l }
    \hline \hline
    State and physical system& $N_{\mathrm{eff}} \geq$ & Using equation \\
    \hline
    Squeezed state, single photonic mode  \cite{Vahlbruch_Detection_2016}  & $31.6$ & (\ref{eq:5})\\
    Squeezed state, two photonic modes  \cite{Eberle_Stable_2013}  & $11.1 \pm 0.3$ & (\ref{eq:5})\\
    Spin-squeezed state, cold atomic ensemble \cite{Hosten_Measurement_2016} & $70.8^{+5.1}_{-4.7}$ &  (\ref{eq:5})\\
    $| 0 \rangle^{\otimes N} + \left| 1 \right\rangle^{\otimes N}, N = 8$, ion trap \cite{Monz_14-Qubit_2011}& $5.0 \pm 0.1$ & (\ref{eq:6}), w/ fitting\\
    One-mode photonic cat state, $\alpha = 2.8$ \cite{Vlastakis_Deterministically_2013}  & $10.2 \pm 0.2$ & (\ref{eq:6}), w/ fitting\\
    Single ion; cat state in spatial mode, $\alpha = 5.9$ \cite{Kienzler_Observation_2016}  & $49.4 \pm 11.6$ & (\ref{eq:6}), w/o fitting\\
    \textit{Idem}& $43.4 \pm 4.3$ & (\ref{eq:6}), w/ fitting\\
    Two-mode photonic cat state, $\alpha = 2.7, \beta =  3.1$ \cite{Wang_Schrodinger_2016}  & $20.0 \pm 2.5$ & (\ref{eq:6}), w/o fitting\\
    \textit{Idem}      & $21.5\pm 2.6$ & (\ref{eq:6}), w/ fitting\\
    \hline
  \end{tabular*}
  \caption{Lower bounds on the effective size for some recent experiments. The remark on fitting refers to the technique applied to obtain the bound (see section \ref{sec:appl-exper}).}
  \label{tab:expbounds}
\end{table*}

We now apply the bounds (\ref{eq:5}) and (\ref{eq:6}) to experiments and discuss certain details. 
For squeezed states, the analysis is implicitly done in most of the published works. The right hand side of equation (\ref{eq:5}) is the inverse of the so-called squeezing parameter \cite{Sorensen_Many-particle_2001} and is explicitly given in many papers. The reader is referred to table \ref{tab:expbounds} where examples of photonic and atomic squeezing are given. Many experiments nowadays overcome 10 dB of squeezing, implying an effective size $N_{\mathrm{eff}} \geq 10$. A recent highlight is a spin-squeezing experiment with half a million of cold rubidium atoms \cite{Hosten_Measurement_2016}, where an effective size of 70 was achieved. 

A bit more work has to be done for cat states. As mentioned before, a (partial) Wigner tomography is a typical way to characterize these states. In the following, we discuss different levels of assumptions or post-processing that lead to different results when using equation (\ref{eq:6}). For illustrations, we use the data for the cat state generated in the experiment of reference \cite{Kienzler_Observation_2016} with $\alpha = 5.9$ [see figure \ref{fig:home} (a) for the data and the fit to equation (\ref{eq:7})].

We start with a direct application of bound (\ref{eq:6}). Experimentally, the data for $W(0,\theta)$ is taken for a discrete set of values $\theta = n \theta_0$, for $ n\in \mathbbm{N}$ and some fixed $\theta_0$. In principle, every pair $p_{\pm} = 1/2[1\pm W(0,n\theta_0)]$ and $q_{\pm} = 1/2[1\pm W(0,m\theta_0)]$ serves as a basis for the bound. However, only values $|n - m|\theta_0 \sqrt{\mathcal{I}_{\rho}(X_{\theta})} \ll 1$ lead to tight bounds. In figure \ref{fig:home} (b), bound (\ref{eq:6}) is computed for all pairs with $|n-m|\leq 3$. Nearest-neighbor pairs give rise to the largest bounds. The error bars are propagated from the error bars of the data points. 

A second way to estimate the QFI from below is to work with a fit of the data. For this, we assume that the dominant noise is an amplitude damping of the coherence term, that is, $\left| \alpha \right\rangle\!\left\langle -\alpha\right| \rightarrow A \left| \alpha \right\rangle\!\left\langle -\alpha\right|$, with $0\leq A<1$ appearing already in equation (\ref{eq:7}). By defining $p_{\pm}$ and $q_{\pm}$ for a pair $(\theta_1,\theta_2)$ similarly as before, a valid bound on the QFI can be computed. The only thing left is to optimize over the real pair $(\theta_1,\theta_2)$, which gives some numerical value (see table \ref{tab:expbounds}). The error bars for the bound are calculated from the uncertainties of the data fitting. A simple formula is found for the special case of $\phi = \pi/2$, which reads
\begin{equation}
\mathcal{I}_{\mathrm{cat}}(x) \gtrsim A^2 S\label{eq:8}
\end{equation}
for $\theta_1 = 0$ and $\theta_2 \rightarrow 0$. 
Equation (\ref{eq:8}) is useful for a quick judgment of the experimental achievements (even if $\phi \neq \pi/2$).
For example, the results $S = 11.8$ and $A = 0.44$ for the cat state reported in reference \cite{Deleglise_Reconstruction_2008} roughly imply $N_{\mathrm{eff}} \approx 2.3$. Further examples are given in table \ref{tab:expbounds}. The advantage of working with fits instead of the direct data are simple expressions like equation (\ref{eq:8}) and smoothing over statistical fluctuations. The drawback is the additional model that has to be justified.

A similar calculation can be done for the corresponding protocol in atomic ensembles which leads basically to the same formula $N_{\mathrm{eff}} \gtrsim A^2 N$. With the data measured in reference \cite{Monz_14-Qubit_2011}, one finds $N_{\mathrm{eff}} \gtrsim 5$ for $N = 8$ (see also \cite{Pezze_Witnessing_2015}).

In many cases, the parity measurement is a processed quantity from a Fock basis measurement. However, the probabilities $f_n = \left\langle n \right| \rho \left| n \right\rangle $ and $g_n = \left\langle n \right| U\rho U^{\dagger} \left| n \right\rangle, n\in \mathbbm{N}_0 $ inferred from the Fock basis measurement can be directly used for bound (\ref{eq:6}). The question is then whether it is preferable to work with $f_n,g_n$ rather than with the Wigner function. As mentioned before, the bound using the Wigner function cut (\ref{eq:7}) is almost tight. Therefore, using a different set of probabilities cannot help a lot. The example from reference \cite{Kienzler_Observation_2016} gives us further insight to resolve this question. The problem with $f_n,g_n$ are the statistical fluctuations, which can be much more pronounced than the difference $f_n-g_n$. This leads to bounds that have a propagated error much larger than the bound itself, that is, the bound has no significance. Working with the parity probabilities has the advantage that the fluctuations are reduced by averaging them.

The present analysis can be extended to many modes. In reference \cite{Wang_Schrodinger_2016}, a two-mode cat state $| \alpha,\beta \rangle + \left| -\alpha,-\beta \right\rangle $ was generated. The optimal generator reads $x^{(1)}+x^{(2)}$, given that $\alpha,\beta \in \mathbbm{R}$. The results using bound (\ref{eq:6}) without and with fitting are given in table \ref{tab:expbounds}.

\section{Discussion and conclusion}
\label{sec:disc-concl}

To summarize, there exist many setups with measurements that allow one to bound the QFI from below. 
In this paper, we put the emphasis on interpreting the QFI as a measure for the extent of quantum coherence in a given spectrum. Even if true macroscopic sizes are currently out of reach, impressive results have been produced so far. With the help of the QFI, different state classes are comparable. For example, squeezed states and cat states can both show large quantum coherence, besides their differences (e.g., negativity of the Wigner function). Nevertheless, one should be careful with judgments like ``experiment $x$ is more successful than experiment $y$'' because of a larger QFI. Many open questions regarding the use of the QFI as a measure for macroscopic quantum states still need to be resolved. Note that also other characterizations of (generalized) Schr\"odinger-cat states like \cite{Lee_Quantification_2011} can be measured in the lab \cite{Jeong_Detecting_2014}.

Besides squeezed and cat states, Fock states and Dicke states (see tables \ref{tab:purePS} and \ref{tab:pureN}) seem to be promising state classes for large quantum coherence. For example, experiments like \cite{Lucke_Detecting_2014,Harder_Source_2015} could yield a large $N_{\mathrm{eff}}$ given the measurements necessary to use bounds (\ref{eq:5}) and (\ref{eq:6}) are feasible. An alternative approach to bound the QFI from below was presented in \cite{Apellaniz_Optimal_2015}, which was applied to Dicke-like states \cite{Lucke_Detecting_2014}. There a lower bound $\mathcal{I} \gtrsim 2.94$ was estimated from the experimental data.

Here, we considered the relative QFI, that is, the QFI was divided by the QFI of a predefined classical state. Another way is to take the square root of the QFI as a measure for coherence length. Again, let us consider the single ion that was brought into a cat state in a spatial degree of freedom \cite{Kienzler_Observation_2016}. There, $\Delta\alpha = 2 \alpha$ is related to the spread in position space. As mentioned in reference \cite{Kienzler_Observation_2016}, $\Delta\alpha = 15.9$ corresponds to a distance of up to $240 \mathrm{nm}$. Taking $\alpha = 5.9$ and the finite amplitude $A = 0.57$, the correlation length is reduced to  $\sqrt{I(x)} \approx \Delta\alpha A \approx 92 \mathrm{nm}$.
For this specific system, one might ask whether it makes sense to talk about macroscopic quantum states, since, after all, one considers only a single ion. To counter this argument, we take into account the harmonic potential of the ion trap. Large $\alpha$ implies large energies $\hbar\omega(|\alpha|^2+1/2)$. This is comparable with photonic systems, where ``large'' refers to the number of photons (i.e., field excitations).

Speculations about achievable values for the QFI and hence for the effective size in the future are difficult. The main problem is the sensitivity of states with large $\mathcal{I}(X)$ to all sources of noise that are generated by $X$. Some problems, like finite measurement precision, can in principle be overcome \cite{Dunningham_Interferometry_2002,Leibfried_Heisenberg-Limited_2004,Wang_Precision_2013,Davis_Approaching_2016,Frowis_Detecting_2016}. Recently, there was a proposal to counter collective phase noise that might be applied in some situations \cite{Landini_Phase-noise_2014}. Finally, stronger lower bounds as presented in \cite{Apellaniz_Optimal_2015} could lead to improved results. Together with further technological progress, numbers much larger than presented in this paper are reachable. Whether this suffices to enter ``true'' macroscopic regimes is nevertheless open.

\section*{Acknowledgments}
\label{sec:acknowledgments}

I am grateful for useful discussions with Nicolas Gisin and Benjamin Yadin and for access to experimental data from the groups of references \cite{Monz_14-Qubit_2011,Vlastakis_Deterministically_2013,Wang_Schrodinger_2016,Kienzler_Observation_2016}. This work was supported by the National Swiss Science Foundation (SNSF) project number 200021\_149109 and the
European Research Council (ERC MEC).

\bibliographystyle{apsrev4-1}
\bibliography{ExperimentalBoundsQFI}
\end{document}